\documentclass[english]{elsart}

\usepackage{amssymb}
\newtheorem{theorem}{\sc Theorem}

\newtheorem{lemma}{\sc Lemma}
\newtheorem{coro}{\sc Corollary}
\newtheorem{req}{\sc Requirement}
\newtheorem{nota}{\sc Notation}
\newtheorem{defin}{\sc Definition}
\newtheorem{ex}{\sc Example}

\newenvironment{proof}{\par \sf Proof.\rm}{\hspace*{\fill}$\Box$\vspace{1ex}}
\newenvironment{corollary}{\begin{coro}}{\end{coro}}

\renewcommand{\emptyset}{\varnothing}

\begin{document}

\begin{frontmatter}



\title{{\Large \bf Time-Bounded Incompressibility of Compressible Strings and Sequences}
}


\author[ILLC]{Edgar G. Daylight\thanksref{kvo}}
\ead{egdaylight@yahoo.com}
\author[CWI]{Wouter M. Koolen}
\ead{W.M.Koolen-Wijkstra@cwi.nl}
\author[CWI,UvA]{Paul M.B. Vit\'anyi}
\ead{Paul.Vitanyi@cwi.nl}

\thanks[kvo]{a.k.a. Karel van Oudheusden}

\address[ILLC]{University of Amsterdam, Institute of Logic, Language, and Computation,Amsterdam, The Netherlands}
\address[CWI]{Centrum voor Wiskunde en Informatica, Science Park 123, 1098 XG Amsterdam, The Netherlands}
\address[UvA]{University of Amsterdam, Department of Computer Science, Amsterdam, The Netherlands}

\begin{abstract}
For every total recursive
time bound $t$, a constant fraction of all 
compressible (low Kolmogorov complexity) strings is $t$-bounded
incompressible (high time-bounded Kolmogorov complexity); 
there are uncountably many infinite sequences of which
every initial segment of length $n$ is compressible to $\log n$ 
yet $t$-bounded incompressible below $\frac{1}{4}n - \log n$;
and there are a countably infinite  number of recursive infinite sequences
of which
every initial segment is similarly $t$-bounded incompressible.
These results and their proofs are related to, 
but different from, Barzdins's lemma.
\end{abstract}

\end{frontmatter}


\section{\label{sec:Introduction}Introduction}

Informally, the Kolmogorov complexity of a finite binary string
is the length of the shortest string from which the original
can be losslessly reconstructed by an effective 
general-purpose computer such as a particular universal Turing machine $U$.
Hence it constitutes a lower bound on how far a 
lossless compression program can compress.
Formally, the {\em conditional Kolmogorov complexity} 
$C(x|y)$ is the length of the shortest input $z$
such that the universal Turing machine $U$ on input $z$ with
auxiliary information $y$ outputs $x$. The 
{\em unconditional Kolmogorov complexity} $C(x)$ is defined by 
$C(x|\epsilon)$ where $\epsilon$ is the empty string (of length 0).
Let $t$ be a total recursive function.
Then, the {\em time-bounded conditional Kolmogorov complexity}
$C^t(x|y)$ is the length of the shortest input $z$
such that the universal Turing machine $U$ on input $z$ with
auxiliary information $y$ outputs $x$ within $t(n)$ steps
where $n$ is the length in bits of $x$.
The {\em time-bounded unconditional Kolmogorov complexity} $C^t(x)$
is defined by $C^t(x|\epsilon)$. 
For an introduction to the definitions and notions
of Kolmogorov complexity (algorithmic information theory)
see ~\cite{LiVi08}.

\subsection{Related Work}\label{sect.related}
Already in 1968 J. Barzdins~\cite{Barzdins1968} obtained
a result known as {\em Barzdins's lemma}, probably the first
result in resource-bounded Kolmogorov complexity, of which the
lemma below quotes the items that are relevant here. Let
$\chi$ denote the characteristic sequence of an arbitrary recursively
enumerable (r.e.) subset $A$ of the natural numbers. That is, $\chi$ is an infinite sequence
$\chi_{1}\chi_{2}\ldots$ where bit $\chi_{i}$ equals $1$ if and
only if $i\in A$. Let $\chi_{1:n}$ denote the first $n$ bits of
$\chi$, and let $C(\chi_{1:n}| n)$ denote the conditional
Kolmogorov complexity of $\chi_{1:n}$, given the number $n$. 

\begin{lemma}\label{lem.barzdins}
{\rm (i)} For every characteristic sequence $\chi$ of a r.e. set $A$
there exists a constant $c$ such that for all $n$ we have
$C(\chi_{1:n}| n)\leq
\log n+ c$.

{\rm (ii)} 
There exists a r.e. set $A$ with characteristic sequence $\chi$
such that for every total recursive function $t$ there is a constant
$c_t$ with $0<c_{t}<1$ such that for all $n$ we have
$C^{t}(\chi_{1:n}| n)\geq c_{t} n$.
\end{lemma}

Barzdins actually proved this statement
in terms of D.W. Loveland's version of Kolmogorov complexity
\cite{Lo69}, which is a slightly different setting. He also proved that there
is a r.e. set such that its characteristic sequence 
$\chi=\chi_{1}\chi_{2}\ldots $ satisfies 
$C(\chi_{1:n}) \geq \log n$ for every $n$.
Kummer \cite{Ku96}, Theorem 3.1,
solving the open problem in Exercise 2.59 of the first edition
of \cite{LiVi08} proved that there exists a r.e. set
such that its characteristic sequence $\zeta=\zeta_1 ,\zeta_2, \ldots$
satisfies $C(\zeta_{1:n}) \geq 2 \log n -c$ for some constant $c$ and
infinitely many $n$.  

The converse of item (i) does not
hold. To see this, consider a sequence $\chi=
\chi_{1}\chi_{2}\ldots $ and 
a constant $c' \geq 2$, such that
for every $n$  we have $C(\chi_{1:n}| n)\geq n-c' \log n$ 
By  item (i), $\chi$
can not be the characteristic sequence of a r.e. set. Transform $\chi$
into a new sequence $\zeta=\chi_{1}\alpha_{1}\chi_{2}\alpha_{2}\ldots$
with $\alpha_{i}=0^{2^{i}}$, a string of $0$s of length $2^{i}$.
While obviously  $\zeta$ can not be the characteristic 
sequence of a r.e. set, there is a constant $c$ such that for every $n$
we have that $C(\zeta_{1:n}| n)\leq \log n+ c$. 

Item (i) is easy to prove and item (ii) is hard to prove.
Putting items (i) and (ii) together, there is a characteristic
sequence $\chi$ of a r.e. set $A$ whose initial
segments are both logarithmic compressible and
time-bounded linearly incompressible, for every
total recursive time bound.
Below, we identify the natural numbers with finite binary strings
according to the pairing 
$
( \epsilon , 0), (0,1), (1,2), (00,3), (01,4), \ldots ,
$
where $\epsilon$ again denotes the empty string.

\subsection{Present Results}

\begin{theorem}
Let $k_0, k_1$ be positive integer constants and
$t$ a total recursive function.

{\rm (i)} 
A constant fraction of all strings $x$ 
of length $n$ with $C(x|n) \leq k_0 \log n$ satisfies 
$C^{t}(x| n)\geq n-k_{1}$.
{\rm (}Lemma~\ref{thm:tincompressible}{\rm )}.

{\rm (ii)} Let 
$t(n) \geq cn$ for $c > 1$ sufficiently large.
A constant fraction of all strings $x$ 
of length $n$ with $C(x|n) \leq k_0 \log n$ satisfies
$C^{t}(x| n) \leq k_0 \log n$
{\rm (}Lemma~\ref{thm:tcompressible}{\rm )}.

{\rm (iii)}  
There exist uncountably many {\rm (}actually $2^{\aleph_0}${\rm )} 
infinite binary sequences $\omega$ such
that $C(\omega_{1:n}|n) \leq \log n$ and 
$C^t(\omega_{1:n}|n) \geq \frac{1}{4} n-\log n$ for every $n$; moreover,
there exist a countably infinite  number of {\rm (}that is $\aleph_0${\rm )} 
recursive infinite binary sequences $\omega$ 
{\rm (}hence $C(\omega_{1:n}|n)=O(1)${\rm )} such that
$C^t(\omega_{1:n}|n) \geq \frac{1}{4} n-\log n$ for every $n$
{\rm (}Lemma~\ref{lem.infinite}{\rm )}.
\end{theorem}
Note that the order of quantification in 
Barzdins's lemma is ``there exists a r.e. set
such that for every total recursive function $t$ 
there exists a constant $c_t$.'' In contrast, in item (iii) 
we prove 
``there is a positive constant
such that for every total 
recursive function $t$ there is a sequence $\omega$.''
While Barzdins's lemma proves the existence of a single characteristic 
sequence of a r.e. set that is time-limited linearly incompressible,
in item (iii) we prove the existence of
uncountably many sequences that are logarithmically compressible
over the initial segments, and the existence of a countably infinite  number of recursive sequences,
such that all those sequences are time-limited linearly incompressible.

We generalize item (i) in Corollaries~\ref{cor:cortincompressible} and
\ref{cor.general}.
Section~\ref{sec:Preliminaries}
presents preliminaries. Section~\ref{sec:Main-Result} gives
the results on finite strings.
Section~\ref{sect.infinite}
gives the results on infinite sequences. Finally, conclusions are presented
in Section~\ref{sec:Conclusions}.
The proofs for the results are different from Barzdins's proofs.

\section{\label{sec:Preliminaries}Preliminaries}

A (binary) program is a concatenation of instructions, and an instruction
is merely a string. Hence, we may view a program as a string.
A program and a Turing machine (or machine for short) are used
synonymously. The length in bits of a string $x$ is
denoted by $|x|$. If $m$ is a natural number, then $|m|$ is
the length in bits of the $m$th binary string in length-increasing
lexicographic order, starting with the empty string $\epsilon$.
We
also use the notation $|S|$ to denote the cardinality
of a set $S$.

Consider a standard enumeration of all Turing
machines $T_{1}$, $T_{2}$, $\ldots .$ Let $U$ denote a universal
Turing machine such that for every $y\in\{0,1\}^{*}$ and $i\geq1$
we have $U(i,y)=T_{i}(y)$. That is, for all finite binary strings
$y$ and every machine index $i\geq1$, we have that $U$'s execution
on inputs $i$ and $y$ results in the same output as that obtained
by executing $T_{i}$ on input $y$. 
Let $t$ be a total recursive function.
Fix $U$ and define
that $C(x| y)$ equals  
$\min\{ |p|: p\in\{0,1\}^{*}\:\textrm{and}\: U(p,y)=x\}$. 
For the same fixed $U$, define that
$C^{t}(x| y)$ equals $\min\{ |p|:\, p\in\{0,1\}^{*}\:\textrm{and}\; U(p,y)=x\;\textrm{in $t(|x|)$ steps}\} $. 
(By definition the sets over which is minimized 
are countable and not empty).

\section{Finite Strings}\label{sec:Main-Result}

\begin{lemma}
\label{thm:tincompressible}
Let $k_0,k_1$ be positive integer constants and $t$ be a total
recursive function.
There is a positive constant $c_t$
such that for sufficiently large $n$ the strings
$x$ of length $n$ satisfying 
$C^{t}(x| n)\geq n-k_{1}$ form
a $c_t$-fraction of the strings $y$
of length $n$ satisfying $C(y|n) \leq k_0 \log n$.
\end{lemma}
\begin{proof}
The proof is by diagonalization.
We use the following algorithm with inputs $t,n,k_1$ and a
natural number $m$.

{\bf Algorithm} ${\cal A}(t,n,k_1,m)$

{\bf Step 1.} Using the universal reference Turing machine $U$,
recursively
enumerate a finite
list of all binary programs $p$ of length $|p|<
n-k_{1}$. There are at most $2^{n}/2^{k_{1}}-1$
such programs. Execute each of these programs on
input $n$. Consider the set of all
programs that halt within $t(n)$ steps and which output
precisely $n$ bits. Call the set of these outputs $B$.
Note that $|B| \leq 2^{n}/2^{k_{1}}-1$ and it 
can be computed in time $O(2^{n} t(n)/2^{k_{1}})$. 

{\bf Step 2.} Output the $(m+1)$th
string of length $n$, say $x$, in the lexicographic order of all strings
in $\{0,1\}^n \setminus B$
and halt. 
If there is no such
string then halt with output $\perp$.
{\bf End of Algorithm}

Because of the selection
process in Step 1, $|\{0,1\}^n \setminus B| \geq 2^n - 2^n/2^{k_1}+1$
and every $x \in \{0,1\}^n \setminus B$ has
time-bounded complexity 
\begin{equation}\label{eq.gt}
C^t(x|n) \geq n-k_1.
\end{equation}
For $|m| \leq k_0 \log n -c$, where the constant $c$ is defined below,
and provided 
$\{0,1\}^n \setminus B$ is sufficiently large, 
that is,
\begin{equation}\label{eq.eq}
n^{k_0}/2^c \leq 2^n \left(1-\frac{1}{2^{k_{1}}}\right)+1,
\end{equation}
there are at least $n^{k_0}/2^c$ strings $x$ of length $n$ 
that will be output by the algorithm. Call this set $D$.
Each string $x \in D$ satisfies 
\begin{equation}\label{eq.lt}
C(x|t,n,k_1,{\cal A},p) \leq |m| \leq k_0 \log n -c.
\end{equation}
Since we can describe the fixed $t,k_0,k_1,{\cal A}$, a program $p$ to
reconstruct $x$ from these data, and the means to tell them
apart, in an additional constant
number of bits, say $c$ bits (in this way the quantity $c$ can be
deduced from the conditional),
it follows that 
$
C(x|n) \leq k_0 \log n
$.
For given $k_0,k_1$, and $c$, inequality (\ref{eq.eq}) 
holds for every sufficiently large $n$. For such sufficiently large $n$, 
the cardinality of the set of strings of length $n$ satisfying
both $C(x|n) \leq k_0 \log n$ and $C^t(x|n) \geq n-k_1$ is at least
$
|D| = n^{k_0}/2^c.
$
Since the number of strings $x$ of length $n$ 
satisfying $C(x|n) \leq k_0 \log n$ is at most
$\sum_{i=0}^{k_0 \log n} 2^i < 2n^{k_0}$, 
the lemma follows with
$c_t = 1/2^{c+1}$.
\end{proof}

\begin{corollary}
\label{cor:cortincompressible}
\rm
Let $k_0$ be a positive integer constant and $t$ be a total recursive function.
For every sufficiently large natural number $n$, 
the set of strings $x$ of
length $n$ such that $C^t(x|n) \not\leq k_0 \log n$ is a positive 
constant fraction of the strings $y$ of length $n$ satisfying 
$C(y|n) \leq k_0 \log n$.
\end{corollary}

We can generalize Lemma~\ref{thm:tincompressible}. 
Let  $t$ be a total recursive function, and $f,g$ 
be total recursive functions such that (\ref{eq.functions})
below is satisfied. 
\begin{corollary}\label{cor.general}
\rm
For every 
sufficiently large natural number $n$, the set of strings $x$ of length $n$
that satisfy both $C(x|n) \leq f(n)$ and $C^t(x|n) \geq g(n)$ is
a positive constant fraction of the strings $y$ of length $n$
satisfying $C(y|n) \leq f(n)$.
\end{corollary}
\begin{proof}
Use a similar algorithm ${\cal A}(t,n,g,m)$ with $|p| <
g(n)$ in Step 1, and $|m| \leq f(n)- c$ in the analysis.
Require
\begin{equation}\label{eq.functions}
2^{f(n)-c} \leq 2^n - 2^{g(n)}+1.
\end{equation}
\end{proof}

\begin{lemma}
\label{thm:tcompressible}
Let $t$ be a total recursive function with $t(n) \geq cn$ for some $c > 1$ 
and $k_0$ be a positive integer constant. For every sufficiently large
 natural number $n$,
there is a positive constant
$c_t$ such that the set of strings $x$ of length $n$
satisfying $C^t(x|n) \leq k_0 \log n$ is a $c_t$-fraction
of the set of strings $y$ of length $n$ satisfying 
$C(y|n) \leq k_0 \log n$.
\end{lemma}
\begin{proof}
We use the following algorithm that takes positive integers $n,m$
as inputs and computes a 
string $x$ of length $n$ satisfying
$C^t(x|n) \leq k_0 \log n -c$.

{\bf Algorithm} ${\cal B}(n,m)$

Output the string $0^{n-|m+1|} (m +1)$ (where $|m+1|$ is the length of 
the string representation of $m+1$) and halt.
{\bf End of Algorithm}

Let $k_0$ be a postive integer and $c$ a positive integer constant 
chosen below.
Consider strings $x$ that are output by algorithm ${\cal B}$ and that
 satisfy
$C^t(x|n,{\cal B},p) \leq |m| \leq k_0  \log n -c$
with $c$ the number of bits to contain 
descriptions of ${\cal B}$ and $k_0$, a program $p$
to reconstruct $x$ from these data, and the means to
tell the  constituent items apart. Hence, $C^t(x|n) \leq k_0 \log n$.
The running time of algorithm ${\cal B}$ is $t(n) = O(n)$, since
the output strings are length $n$ and to output the $m$th string with
$m \leq 2^{k_0 \log n -c}$ we simply take the binary
representation of $m$ and pad it with nonsignificant 0s to
length $n$.
Obviously, the strings that satisfy
$C^t(x|n)\leq k_0 \log n$ are a subset of the strings that satisfy
$C(x|n)\leq k_0 \log n$. 
 There are at least $n^{k_0}/2^c$  strings of the first kind
while there are at most $2n^{k_0}$ strings of the second kind.
Setting $c_t = 1/2^{c+1}$ finishes the proof.
\end{proof}

It is well known that if we flip a fair coin $n$ times, that is,
given $n$ random bits, then we obtain
a string $x$ of length $n$ with Kolmogorov complexity $C(x|n) \geq n-c$
with probability at least $1- 2^{-c}$. Such a string $x$ is
algorithmically random. We can also get by with less random bits to
obtain resource-bounded algorithmic randomness from compressible strings.
\begin{lemma} 
Let $a,b$ be constants as in the proof below.
Given the set of strings $x$ of length $n$
satisfying $C(x|n) \leq k_0 \log n$, a total recursive function $t$, 
the constant $k_1$
as before, and $O(ab \log n)$ fair coin 
flips, we obtain
a set of $O(ab)$ strings of length $n$ such that with
probability at least $1-1/2^b$ one string $x$ in this set
satisfies $C^t(x|n) \geq n-k_1$.
\end{lemma}
\begin{proof}
By Lemma~\ref{thm:tincompressible}, a $c_t$th fraction of the set $A$
of strings
$x$ of length $n$ that have 
$C(x|n) \leq k_0 \log n$ also have $C^t(x|n) \geq n-k_1$.
Therefore, by choosing,
uniformly at random, a constant
number $a$ of strings from the set $A$
we increase (e.g. by means
of a Chernoff bound \cite{LiVi08}) the probability 
that (at least) one of those strings cannot be compressed below $n-k_1$
in time $t(n)$
to at least $\frac{1}{2}$. To choose any one string from $A$ 
requires $O(\log n)$ random bits by dividing
$A$ in two equal size parts and repeating this with
the chosen half, and so on. The selected $a$ elements take
$O(a \log n)$ random bits. Applying the
previous step $b$ times, the probability
that at least one of the $ab$ chosen strings cannot be compressed
below $n-k_1$ bits in time $t(n)$
is at least $1- 1/2^{b}$.
\end{proof}

\section{From Finite Strings to Infinite Sequences}\label{sect.infinite}

We prove a result reminiscent of Barzdins's lemma, Lemma~\ref{lem.barzdins}.
In Barzdins's version, 
characteristic sequences $\omega$ of r.e. sets are considered
which by Lemma~\ref{lem.barzdins} have complexity
$C(\omega_{1:n}|n) \leq \log n +c$. Here, we consider 
a wider class of sequences of which the initial segments
are logarithmically compressible (such sequences are not necessarily
characteristic sequences of r.e. sets as 
explained in Section~\ref{sect.related}).

\begin{lemma}\label{lem.infinite}
Let $t$ be a total recursive function.
{\rm (i)} There are uncountably many {\rm (}actually 
$2^{\aleph_0}${\rm )} sequences 
$\omega = \omega_1 \omega_2 \ldots$ 
such that both $C(\omega_{1:n}| n) \leq \log n$ and 
$C^{t}(\omega_{1:n}| n) \geq \frac{1}{4}n-\log n$ for every $n$.

{\rm (ii)} The  set in item {\rm (i)}
contains a countably infinite  number of {\rm (}that is $\aleph_0${\rm )} recursive sequences
 $\omega = \omega_1 \omega_2 \ldots$
such that $C^{t}(\omega_{1:n}| n) \geq \frac{1}{4}n-\log n$ for every $n$. 
\end{lemma}
\begin{proof}
(i) Let $g(n)=\frac{1}{2}n- \log n$. 
Let $c \geq 2$ be a constant to be chosen later, $m_i = c2^i$,
$B(i),C(i),D(i) \subseteq \{0,1\}^{m_i}$ 
for $i=0,1, \ldots$, and $C(-1)= \{\epsilon\}$. The $C$ sets are constructed
so that they contain the target
strings in the form of a binary tree, where $C(i)$ contains all target strings
of length $m_i$. 
The $B(i)$ sets correspond to forbidden prefixes of length $m_i$. The $D(i)$ sets  
consist of the set of strings of length $m_i$ with prefixes
in $C(i-1)$ from which the strings in $C(i)$ are selected.

{\bf Algorithm} ${\cal C}(t,g)$:

{\bf for} $i :=0,1, \ldots $ {\bf do}

{\bf Step 1.} Using the universal reference Turing machine $U$,
recursively
enumerate the finite
list of all binary programs $p$ of length $|p|<
g(m_i)$ with $m_i = c2^i$ and the constant $c$
defined below. There are at most $2^{g(m_i)}-1$
such programs. Execute each of these programs on
all inputs $m_i+j$ with $0 \leq j < m_i$. Consider the set of all
programs  with input $m_i+j$ that halt with output $x=yz$  
within $t(|x|)$ time with $|x|=m_i+j$,
$y \in C(i-1)$ (then $|y| = m_{i-1}$ for $i > 0$ and 
$|y|=0$ for $i=0$), and
$z$ is a binary string such that $x$ satisfies $m_{i} \leq |x| < m_{i+1}$. 
There are at most $m_i (2^{g(m_i)}-1)$ such $x$'s.
Let $B(i)$ be the set of the $m_i$-length prefixes of these $x$'s.
Then, $|B(i)| \leq m_i (2^{g(m_i)}-1)$ and it
can be computed in time $O(m_i2^{g(m_i)} t(m_{i+1}))$.
Note that if $u \in \{0,1\}^{m_i} \setminus B(i)$ then $C^t(uw| \; |uw|)
\geq g(|u|)$ for every $w$ such that $|uw| < m_{i+1}$.

{\bf Step 2.} 
Let $C(i-1)=\{x_1,x_2, \ldots ,x_h\}$ and 
$D(i)= (C(i-1)\{0,1\}^* \bigcap  \{0,1\}^{m_{i}}) \setminus B(i)$.
{\bf for} $l:=1, \ldots ,h$ {\bf do}
{\bf for} $k:=0,1$ {\bf do} put the $k$th string
with initial segment $x_l$, in the 
lexicographic order of $D(i)$,
in $C(i)$.
If there is no such
string then halt with output $\perp$. {\bf od} {\bf od} {\bf od}
{\bf End of Algorithm}

Clearly, $C(i) \{0,1\}^* \subseteq C(i-1)\{0,1\}^*$
for every $i=0,1, \ldots .$
Therefore, if 
\begin{equation}\label{eq.inter}
\bigcap_{i=0}^{\infty} C(i) \{0,1\}^{\infty} \neq \emptyset,
\end{equation}
then the elements of this intersection constitute 
the infinite sequences $\omega$ in the statement of the lemma.

\begin{claim}\label{claim}
\rm
With $g(m_i) = \frac{1}{2} m_i - \log m_i$, we have
$|C(i)| = 2^{i+1}$ for $i=0,1, \ldots .$.
\end{claim}
\begin{proof}
The proof is by induction. Recall that $m_i=c 2^i$ with
the constant $c \geq 2$.
{\em Base case}: $|C(0)|=2$ since $C(-1) = \{\epsilon\}$ and 
$|D(0)| \geq 2^{m_0} - m_0 (2^{g(m_0)} -1) \geq  2$.  

{\em Induction}: Assume that the lemma is true for every $0 \leq j <i$.
Then, every string in $C(i-1)$ has two extensions in $C(i)$, since
for every string in $C(i-1)$ there are $2^{m_i - m_{i-1}}$
extensions available of which at most $|B(i)| \leq m_i (2^{g(m_i)}-1)$
are forbidden. Namely, $2^{m_i - m_{i-1}} - |B(i)| \geq 
2^{m_i/2} - 2^{g(m_i) + \log m_i} + m_i \geq 2$.
Hence it follows that the binary $k$-choice can always be made
in Step 2 of the algorithm for every $l$. Therefore $|C(i)| = 2^{i+1}$.
\end{proof}

Let a constant $c_1$ account for the constant number of bits
to specify the functions $t,g$, the algorithm ${\cal C}$,
and a reconstruction program that executes the following:
We can specify every initial
$m_i$-length segment of a particular $\omega$ in the set on the
lefthand side of (\ref{eq.inter}) by  
running the algorithm ${\cal C}$ using the data represented by the $c_1$
bits, $m_i$,
and the indexes $k_j\in\{0,1\}$ of the strings in 
$D(j)$ with initial segment in $C(j-1)$, 
$0 \leq j \leq i$, that form a prefix of $\omega$. Therefore,
\[
C(\omega_{1:m_i}|m_i) \leq c_1 +i+1.
\]   
Setting $c=2^{c_1+1}$ yields 
$C(\omega_{1:m_i}|m_i) \leq \log c +i = \log m_i $.
By the choice of $B(i)$ in the algorithm we know that
$C^t(\omega_{1:m_i+j}|m_i+j) \geq g(m_i)$ for every $j$ satisfying
$0 \leq j < m_i$. 
Because $2m_i = m_{i+1}$, for every $n$ satisfying $m_i \leq n < m_{i+1}$,
$C^t (\omega_{1:n}|n) \geq \frac{1}{2}m_i - \log m_i \geq \frac{1}{4}n - \log n$.
Since this holds for every $i=0,1, \ldots ,$
item (i) is proven with 
$C^t(\omega_{1:n}|n) \geq \frac{1}{4}n -\log n$
for every $n$. The number of $\omega$'s concerned equals
the number of paths in an infinite complete binary tree, that is,
$2^{\aleph_0}$.

(ii) This is the same as item (i) except 
that we always take, for example, $k_i=0$ (no binary choice)
in Step 2 of the algorithm. In fact, we can specify an
arbitrary computable 0--1 valued function to choose the $k_i$'s.
There are a countably infinite  number of (that is $\aleph_0$) such functions.
The specification of every such function $\phi$ takes $C(\phi)$ bits.
 Hence we do not have to specify 
the successive $k_i$ bits, and 
$C(\omega_{1:n}|n) = c_1 +1+C(\phi)=O(1)$ with $c_1$ the constant in the
proof of item (i).
Trivially, still $C^t(\omega_{1:m_i+j}|m_i+j) \geq g(m_i)$ 
for every $j$ satisfying
$0 \leq j < m_i$. Since this holds for every $i=0,1, \ldots ,$
item (ii) is proven by item (i).
\end{proof}

\section{\label{sec:Conclusions}Conclusions}

We have proved the items promised in the abstract.
In Lemma~\ref{lem.infinite}
we iterated the proof method of
Lemma~\ref{thm:tincompressible} to prove a result
which is
reminiscent of Barzdins's lemma
\ref{lem.barzdins}, relating compressiblity
and time-bounded incompressiblity of infinite sequences in another manner.
Alternatively, we could have studied space-bounded incompressibility.
It is easily verified  that the results 
also hold when the time-bound
$t$ is replaced by a space bound $s$ and the time-bounded Kolmogorov
complexity is replaced by space-bounded
Kolmogorov complexity.

\section*{Acknowledgement}
We thank the referees for comments, references, pointing out an error
in the original proof of Lemma~\ref{thm:tincompressible} and 
that the argument used there is both independent and 
close to that used
to prove Theorem 3.2 in \cite{AFMV}.


\begin{thebibliography}{1}
\bibitem{AFMV}
L. Antunes, L. Fortnow, D. van Melkebeek, and
N. V. Vinodchandran, Computational depth: Concept and applications,
{\em Theor. Comput. Sci.}, 354:3(2006),
391--404.

\bibitem{Barzdins1968}
Ja.M. Barzdins, Complexity of programs to determine
whether natural numbers not greater than $n$ belong to a recursively
enumerable set, {\em Soviet Math. Dokl.},
9(1968), 1251--1254.

\bibitem{LiVi08} 
M. Li and P.M.B. Vit\'anyi, {\em An Introduction
to Kolmogorov Complexity and Its Applications}, Third edition,
 Springer-Verlag,
New York, 2008.

\bibitem{Lo69}
D.W. Loveland, A variant of the Kolmogorov concept of complexity,
{\em Inform. Contr.}, 15(1969), 510-526.

\bibitem{Ku96}
M. Kummer, Kolmogorov complexity and instance complexity of recursively
enumerable sets, {\em SIAM J. Computing}, 25(1996), 1123--1143.

\end{thebibliography}
\end{document}